\documentclass[12pt]{article}
\usepackage{graphicx}
\usepackage{emulateapj}

\def\lapprox{\hbox{\lower .8ex\hbox{$\,\buildrel < \over\sim\,$}}}
\def\gapprox{\hbox{\lower .8ex\hbox{$\,\buildrel > \over\sim\,$}}}

\begin{document}

\title{No surviving companion in Kepler's supernova}

\author{Pilar Ruiz--Lapuente$^{1,2}$, Francesco Damiani$^{3}$, Luigi Bedin$^{4}$,
Jonay I. Gonz\'alez Hern\'andez$^{5,6}$, Llu\'{\i}s Galbany$^{7}$, John 
Pritchard$^{8}$, Ramon Canal$^{2}$, and Javier M\'endez$^{9}$}

\altaffiltext{1}{Instituto de F\'{\i}sica Fundamental, Consejo Superior de 
Investigaciones Cient\'{\i}ficas, c/. Serrano 121, E-28006, Madrid, Spain}
\altaffiltext{2}{Institut de Ci\`encies del Cosmos (UB--IEEC),  c/. Mart\'{\i}
i Franqu\'es 1, E--08028, Barcelona, Spain}

\altaffiltext{3}{INAF, Osservatorio Astronomico di Palermo G.S. Vaiana, Piazza
del Parlamento 1, I-90134, Palermo, Italy}

\altaffiltext{4}{INAF, Osservatorio Astronomico di Padova, Via dell' 
Osservatorio 3, I-35122, Padova, Italy}

\altaffiltext{5}{Instituto de Astrof\'{\i}sica de Canarias, E-38206, La Laguna, 
Tenerife, Spain}

\altaffiltext{6}{Universidad de La Laguna, Departamento de Astrof\'{\i}sica, 
E-38206, La Laguna, Tenerife, Spain}

\altaffiltext{7}{PITT PACC, Department of Physics and Astronomy, University of 
Pittsburg, Pittsburg, PA  15620, USA}

\altaffiltext{8}{European Southern Observatory, Karl-Schwarzschild-Str. 2, 
85748 Garching bei M\"unchen, Germany}

\altaffiltext{9}{Isaac Newton Group of Telescopes, P.O. Box 321, E-38700, 
Santa Cruz de La Palma, Spain}

\begin{abstract}

\noindent
We have surveyed Kepler's supernova remnant in search of the companion star
of the explosion. We have gone as deep as 2.6 $L_{\odot}$ in all stars within 
20\% of the radius of the remnant. We use FLAMES at the VLT-UT2 telescope to
obtain high resolution spectra of the stellar candidates selected from 
{\it HST} images. The resulting set of stellar parameters suggests that these 
stars come from a rather ordinary mixture of field stars (mostly giants). A 
few of the stars seem to have low [Fe/H] ($<$ -1) and they are consistent with 
being metal-poor giants. The radial velocities and rotational velocities 
$v_{rot}$ sin $i$ are very well determined.
There are no fast rotating stars as $v_{rot}$  sin i $<$  20 km 
s$^{-1}$ for all the candidates. The radial velocities from the spectra and 
the proper motions determined from {\it HST} images are compatible with those 
expected from the Besan\c con model of the Galaxy. 
The strong limits placed on luminosity suggest that this 
supernova could have arisen either from the core-degenerate scenario or from 
the double-degenerate scenario.
\end{abstract}


\section{Introduction}

\noindent
The supernova of 1604, observed by Johannes Kepler and other European,
Korean, and Chinese astronomers, is one of the five ``historical'' supernovae 
that have been classified as belonging to the Type Ia (thermonuclear), 
the other four being SN 1572 (Tycho Brahe's SN), SN 1006, SN 185 (supposed to 
have created the remnant RCW86) and the recently discovered youngest SNIa 
G1.9+03 that occurred in our Galaxy as recently as around 1900 but was not 
discovered due to dust exctinction and being only observable from the Southern
hemisphere.

\noindent
As it is well known, Type Ia SN (SNe Ia) are well explained by the 
thermonuclear explosion of a mass--accreting C+O white dwarf star (WD), a 
member of a close binary system, the mass donor being the other component of 
the system. There are three proposed channels to bring the WD to the point of 
explosion, depending on the nature of the companion star. In the 
single--degenerate channel (SD), the companion is a still thermonuclearly 
evolving star (Whelan \& Iben 1973; Nomoto 1982), in the double--degenerate (DD) 
channel it is another WD, either a C+O WD or a He WD (Iben \& Tutukov 1984; 
Webbink 1984). Another possible channel, known as the core--degenerate scenario 
(CD), involves a C+O WD that merges with the core of an Asymptotic Giant 
Branch (AGB) star, following a common--envelope episode (Livio \& Riess 2003; 
Soker, Garc\'{\i}a--Berro \& Althaus 2014; Aznar-Sigu\'an et al. 2015).

\noindent
 Significant progress has been done in the identification of progenitors 
of Type  Ia supernovae (hereafter SNe Ia). For
instance, there has been proof of an specific scenario that works to give rise
to SNe Ia. This is the double detonation scenario studied theoretically by 
Fink et al. (2010); Sim et al (2012) and others.  In this
scenario, the CO WD accumulates a helium--rich layer on in its surface. The
detonation of the helium--rich layer ignites the CO WD. This seems to be the
explosion mechanism involved in MUSSES1604D (Jiang et al. 2017) and similar
events. A He WD companion seems to be favored. The donor He-rich WD
might survive in particular cases studied by Shen \& Schwab (2017).

\noindent
 However, the observed double detonation scenario, as seen from the effect 
in the very early light curves of the SNe Ia, can not account for more than a 
small percentage of the SNe Ia observed (Jiang et al. 2017).

\noindent
 Another possible path to SNe Ia is the WD--WD collision. In this 
new DD path, no surviving companion is expected. A study of this mechanism to
give rise to SNe Ia shows that would account for $<$ 1\% of the observed events
(see Soker 2018 for an overview).

\noindent
So, mainly in the single degenerate scenario, a surviving companion should 
remain after the explosion. There is no surviving companion, but merging of the 
two components of the system in the DD scenario and in the CD scenario. On the 
contrary, in the SD channel, the companion star should survive the SN 
explosion. A surviving companion might be identified from its kinematics 
(large radial velocity and/or proper motion, fast rotation), anomalous 
luminosity, or contamination of its surface layers by the SN ejecta (Wang \& 
Han 2012; Ruiz-Lapuente 2014 review those effects).  Detailed simulations 
can be found in Marietta, Burrows \& Fryxell (2000); Podsiadlowski (2003); 
Pakmor et al. (2008); Pan, Ricker \& Taam (2012,2013,2014); Liu et al.
 (2012, 2013); Shapee, Kochanek \& Stanek (2013); Shen \& Schwab (2017).
 We will compare observations with
their predictions.

\noindent
The central regions of the SNR of Tycho SN (Ruiz-Lapuente et al. 2004; 
Gonz\'alez Hern\'andez et al. 2009; Kerzendorf et al. 2009, 2013; Bedin et al. 
2014) and of SN 1006 (Gonz\'alez Hern\'andez et al. 2012; Kerzendorf et al. 
2012, 2018) have already been explored, to search for a possible surviving 
companion of the SN, as well as extragalactic remnants like SNR 0509-67.5 
(Schaefer \& Pagnotta 2012), SNR 0509-68.7 (Edwards et al. 2013), and N103B 
(Pagnotta \& Schaefer 2015; Li et al. 2017). Studies of other SNRs are in 
progress or have been proposed. Through comparison of the work done by various 
authors in those SNR, the double degenerate scenario seems favored in several 
SNe Ia.

\noindent
The classification of Kepler SN, SN 1604, as a SN Ia was a matter of debate 
for a long time, some authors classified it as  a core--collapse SN, 
in spite of its position, quite above the Galactic plane. The question has 
been settled by X–-ray observations of the remnant (Cassam-Chena\"{\i} et al. 
2004), showing an O/Fe ratio characteristic of SNe Ia (Reynolds et al. 2007).

\noindent
There are indications (Vink 2008) that one component of the binary system 
giving rise to the SN might have created a detached circumstellar shell with a 
mass $\sim$ 1 M$_{\odot}$, expanding into the interstellar medium. More 
recently, Katsuda et al. (2015) have deduced that the shell should have lost 
contact with the binary years before the explosion. It has been suggested 
(Chiotellis et al. 2012; Vink 2016) that the companion star was an AGB star
having lost its envelope at the time of the explosion.

\noindent
The distance to the remnant has also been the object of discussion, the 
estimates ranging between 3 and 7 kpc. Thus, Reynoso \& Goss (1999), based on 
the HI absorption towards the remnant, estimated 4.8 $<$ $d$ $<$ 6.2 kpc. 
Later, Sankrit et al. (2005), from the proper motion of the optical filaments, 
found $d$ = 3.9$^{+1.4}_{-0.9}$ kpc. But very recently, Sankrit et al. 
(2016) have revisited their method and give $d$ = 5.1$^{+0.8}_{-0.7}$ kpc. 
Even more recently, Ruiz-Lapuente (2017), from the reconstruction of the 
optical light curve of the SN based on the historical records, also infers a 
distance $d$ = 5.0$\pm$0.61 kpc, in agreement with Sankrit et al. (2016). We 
thus adopt here a distance d $\sim$ 5.0$\pm$ 0.7 kpc to Kepler SN. For that 
distance, given the Galactic latitude of the SNR, b = 6.8$^{o}$, it lies 
$\simeq$ 590 pc above the Galactic plane.

\noindent
The aim of the paper is to address the progenitor system that led to Kepler's 
supernova, SN 1604. A first paper on the possible progenitor of the Kepler 
supernova suggested a marginal possibility that there was a donor, but only 
tentatively (Kerzendorf et al. 2014, K14 hereafter). At that time, the {\it 
Hubble Space Telescope (HST)} proper motions were not analysed and the stellar 
parameters of the stars were unknown. Here we provide a complete analysis of a 
survey using the FLAMES instrument at the SN ESO VLT-UT2 and we add all the
proper motion information from {\it HST} with a baseline of 10 years.

\noindent
The present paper is organized as follows. Section 2 describes the search in 
Kepler and what can be obtained from it. Section 3 describes the observations 
done with the VLT using the FLAMES instrumentation and the reduction of those 
observations. It presents, as well, the proper motions obtained from data
from the HST archive, from programmes GO-9731 and GO-12885 (P.I: Sankrit).
Section 4 presents the method used to derive the stellar 
parameters and the results. Section 5 presents the estimated distances to the 
stars, and discusses the radial velocities obtained, comparing them to those 
in previous studies. 
Section 6 compares the candidate stars with a kinematical model of the Galaxy. 
Section 7 discusses the results and Section 8 provides a summary of the 
conclusions.

\section{Survey for the progenitor of SN 1604}

\noindent
Our survey has a limiting apparent magnitude $m_{R}$ = +19 mag. The visual 
extinction $A_{V}$, in the direction to the remnant of Kepler's SN is $A_{V}$ = 
2.7$\pm$0.1 mag, and $A_{R}$/$A_{V}$ = 0.748. Thus, we have reached down to an 
absolute magnitude $M_{R}$ = +3.4 mag. That corresponds to a luminosity L = 
2.6 L$_{\odot}$ . For the spectroscopic observations we used FLAMES (Pasquini
et al. 2002) mounted to the UT2 of the VLT. For the measurement of the proper 
motions, we used archival data from the {\it HST}.

\begin{table*}
\scriptsize
       \centering
       \caption{Names and coordinates of the candidate stars}   
       \label{tab:pos}
       \begin{tabular}{lcc}
\\
\hline
\hline

Name & RA (J2000.0) & DC (J2000.0) \\

\hline

T01 & 17 30 39.700 & -21 29 35.54 \\
T02 & 17 30 39.713 & -21 29 46.75 \\ 
T03 & 17 30 40.626 & -21 29 45.02 \\ 
T04 & 17 30 40.161 & -21 29 53.09 \\ 
T05 & 17 30 41.397 & -21 29 44.70 \\ 
T06 & 17 30 40.566 & -21 29 33.80 \\ 
T07 & 17 30 40.732 & -21 29 34.34 \\ 
T08 & 17 30 40.617 & -21 29 27.25 \\ 
T09 & 17 30 40.953 & -21 29 25.56 \\ 
T10 & 17 30 40.897 & -21 29 21.06 \\ 
T11 & 17 30 40.222 & -21 29 18.89 \\ 
T12 & 17 30 40.337 & -21 29 15.16 \\ 
T13 & 17 30 40.707 & -21 29 16.61 \\ 
T14 & 17 30 41.083 & -21 29 14.04 \\ 
T14b& 17 30 41.188 & -21 29 15.32 \\ 
T15 & 17 30 40.985 & -21 29 11.13 \\ 
T16 & 17 30 41.431 & -21 29 06.22 \\ 
T17 & 17 30 39.981 & -21 29 05.86 \\ 
T18 & 17 30 41.823 & -21 29 16.92 \\ 
T19 & 17 30 41.697 & -21 29 25.99 \\ 
T20 & 17 30 42.493 & -21 29 36.22 \\ 
T21 & 17 30 42.653 & -21 29 31.49 \\ 
T22 & 17 30 43.316 & -21 29 24.04 \\ 
T23 & 17 30 43.582 & -21 29 16.96 \\ 
T24 & 17 30 42.651 & -21 29 13.80 \\ 
T25 & 17 30 42.937 & -21 29 52.31 \\ 
T26 & 17 30 42.475 & -21 29 52.90 \\ 
T27 & 17 30 41.898 & -21 30 04.20 \\ 
T28 & 17 30 42.837 & -21 29 45.08 \\ 
T29 & 17 30 41.816 & -21 30 08.89 \\ 
T30 & 17 30 40.335 & -21 30 06.24 \\ 
T31 & 17 30 42.829 & -21 29 02.77 \\ 

\hline

\end{tabular}
\end{table*}

\noindent
As a comparison, Kerzendorf et al. (2014) performed a shallower survey of
possible survivors down to L $>$ 10 L$_{\odot}$, according to them ( L $>$ 6 
L$_{\odot}$, if we take our reliable newly determined distance to SN 1604).

\begin{figure*}
\centering
\includegraphics[angle=-90,width=0.7\paperwidth]{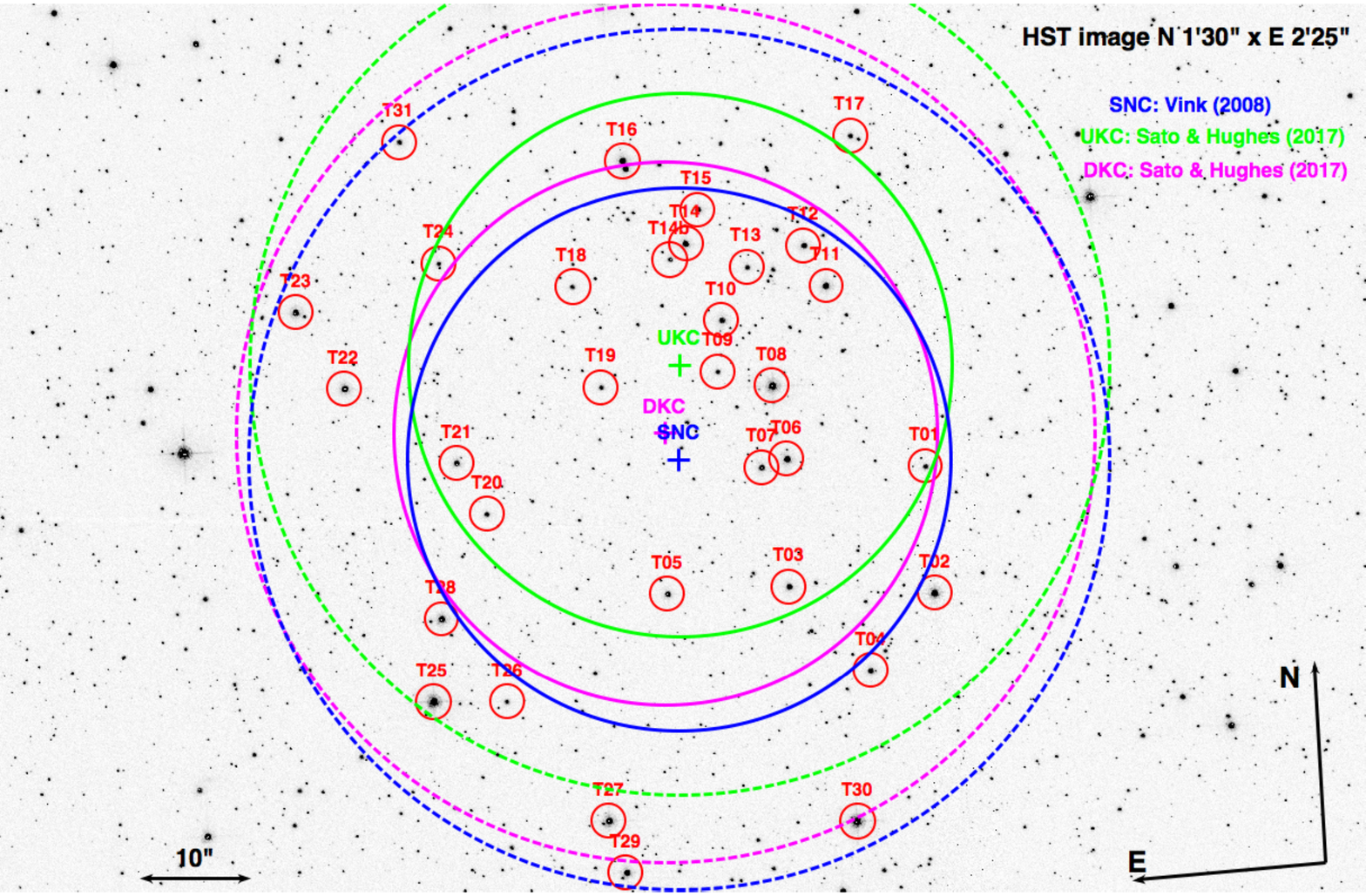}
\includegraphics[angle=-90,width=0.7\paperwidth]{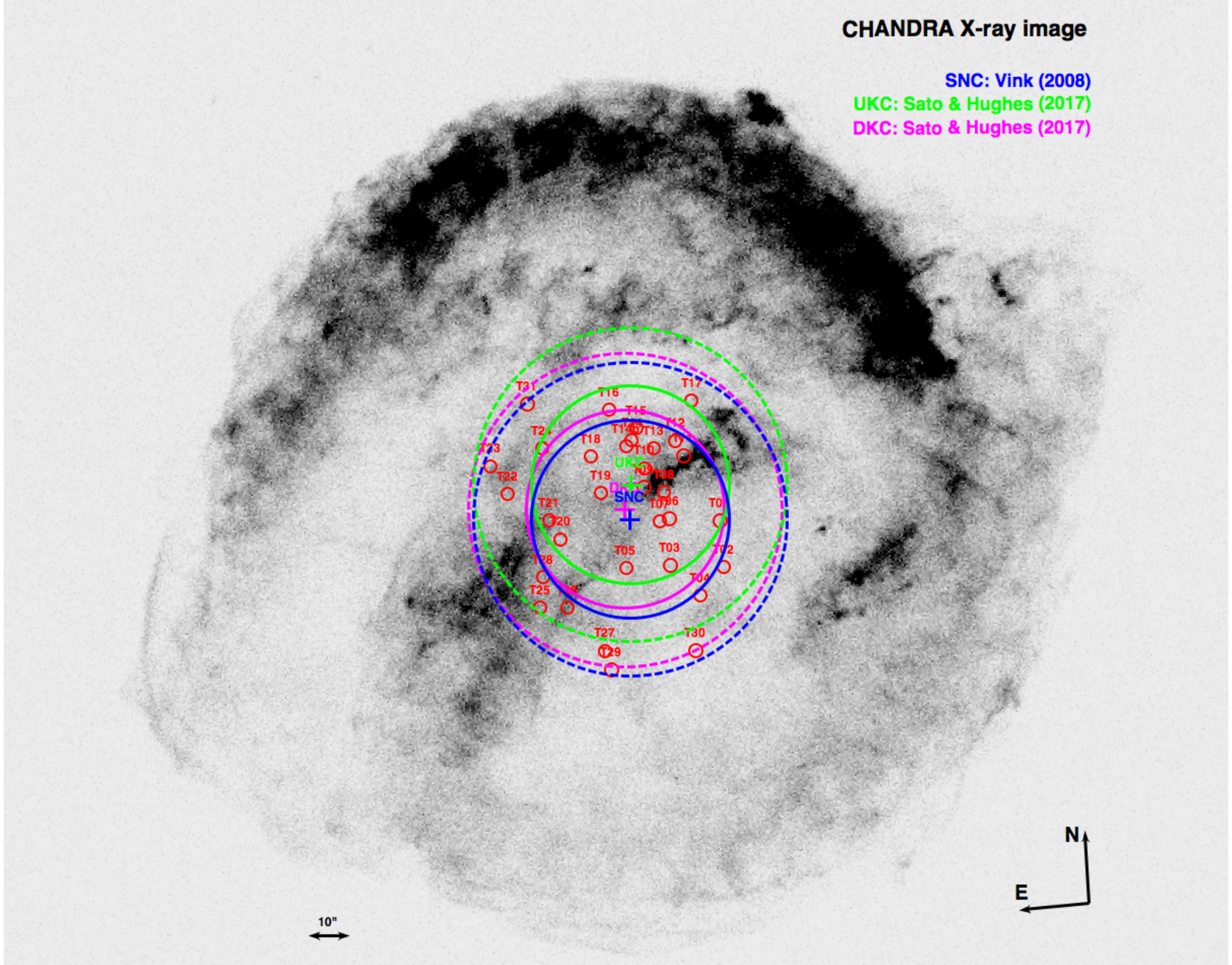}
\caption{\scriptsize 
Figure 1a. 
The targeted stars, in an image from the {\it HST}. Our adopted 
center of Kepler's SNR, at
$\alpha_{J200}$ = 17$^{h}$ 30$^{m}$ 41$^{s}$.25, $\delta_{J2000}$ = -21$^{o}$ 
2'’ 32''.95, is marked with a blue cross. We also provide the undecelerated
 (UKC, green) and decelerated (DKC, magenta) kinematic centers from Sato \& 
Hughes (2017). The big (solid line) blue circle corresponds to a radius of 24
arcsec around the center SNC of the SNR, where our primary targets are located.
The big (dashed) blue circle (of 38 arcsec) around SNC encompasses our 
supplementary targets (see text), and similarly for the circles around UKC
and DKC.
Figure 1b. Chandra X-ray image in the iron-rich 0.7-1.0 keV (Reynolds et al. 
2007; Vink 2008) of the SN 1604, with the regions of search and labels in
Figure 1a. 
}
\end{figure*}

\noindent
The remnant of SN 1604 has an average angular diameter of 225 arcsecs. Our 
survey is complete down to $m_{R}$ $\leq$ 19 within 24 arcsecs of the center 
of the SNR ($\simeq$ 20\% of its radius: blue circumference in 
(Figure 1a), at 
 $\alpha_{J2000}$ = 17$^{h}$ 30$^{m}$ 41$^{s}$.25, $\delta_{J2000}$ = -21$^{o}$ 
29' 32''.95 (Vink 2008). Additional fibers were used to extend the search 
beyond 20\% of its radius (the green circumference encompasses 38 arcsec of 
the radius), although the supplementary stars are not very relevant, due to 
their distances to the center of the SNR. The radius of the search area of 24 
arcsec, at a distance of 5 kpc, corresponds to a transversal displacement from 
the center of the SNR by 0.58 pc. That is the path that a possible companion 
star would have travelled in 400 yr, moving at $v$ = 1460 km s$^{-1}$ 
perpendicularly to the line of sight. A total of 32 stars were observed. They 
are listed, with their coordinates, in Table 1.

\noindent
While preparing the final version of this work, a new analysis of the 
X--ray knots of the Kepler SN by Sato \& Hughes (2017) has provided as a 
result new 
estimates of the expansion center. Both are very close to the center that we 
used, by Vink (2008), thus it does not impact the results of the stars 
included in our search. There is an estimate that does not take into account a 
possible deceleration of the knots. This places the center at $\alpha_{J2000}$ 
= 17$^{h}$ 30$^{m}$ 41$^{s}$.189 $\pm$ 3.6$^{s}$ and $\delta_{J2000}$
= -21$^{o}$ 29' 24''.63 $\pm$ 3.5''. The center that takes into account a 
deceleration coincides practically with that of Vink (2008).
 The newly determined 
center taking into account a model for the deceleration of the knots is 
$\alpha_{J2000}$ = 17$^{h}$ 30$^{m}$ 41$^{s}$.321 $\pm$ 4.4$^{s}$ and 
$\delta_{J2000}$ = -21$^{o}$ 29' 30''.51 $\pm$ 4.3''. We include these new two 
centers in our Figure 1a. We include our search area in relation with the 
whole SNR in Figure 1b.

\noindent
K14  explored the central region of Kepler's SNR. The search has been 
photometric and spectroscopic, covering a 
square field of 38''x 38'' around the center of the SNR determined by Katsuda 
et al. (2008), down to $m_{V} \simeq$ 18 mag. They have used, for their 
spectroscopy, the 2.3m telescope of the ANU, and for the photometry archival 
{\it HST} images. The WiFeS-spectrograph is an image slicer with 25 38X1'' 
slitlets and 0.5'' sampling in the spatial direction on the detector. They 
chose this instrument for its large field of view. However, that 
instrumentation did not allow to determine the stellar parameters. Apparently 
they noted that, due to sky subtraction errors, the continuum placement in 
their data was unreliable. Without determined stellar parameters, it is not
possible to estimate distances, because the absolute magnitudes of the stars 
then remain unknown. They also had problems to estimate rotational velocities 
to better than 200 km s$^{-1}$ , due to the resolution and quality of the 
spectra (K14).

\noindent
 Here, we present a study that includes the stellar parameters, 
T$_{\rm eff}$, log $g$ and [Fe/H]), and we also have rotational velocities and 
radial velocities, apart from the proper motions from {\it HST}. The 
conclusion on the supernova companion is thoroughly tested.

\section{Observations}

\subsection{Spectral observations and reductions}

\noindent
Spectroscopic observations were secured with the multiobject spectrograph 
FLAMES (Pasquini et al. 2002) mounted at the Very Large Telescope (VLT) of the 
European Southern Observatory. Observations were made in the Combined 
IFU/7-Fibre simultaneous calibration UVES mode (Dekker et al. 2000) and 
Giraffe using the HR9 and HR15n settings under ESO programme ID 093.D-0384(A). 
The observations with UVES and Giraffe were done under clear sky and seeing 
conditions ranging from 0.78 to 1.88 arcsec (average of 1.28) from August 3rd 
to 25th 2014. FLAMES is the best instrument to use for our purpose, in
particular Giraffe-IFU, since it provides the possibililty to observe within a 
very small field (24 and 38 arcsec in radius, blue and green circumferences in 
Figure 1) 32 targets to be obtained at the highest possible resolution and to 
minimize the requested observing time. 15 observing blocks (OBs) of 1 hour 
were prepared. Other modes of Giraffe as MEDUSA are not adequate for the 
amount of targets within a small circle of 24 arcsec (see Figure 1a
 as well as the separation between targets).

\noindent
Observations of stars T2, T8, T25, T27, T29, and T30 were carried out with 
UVES using standard settings for the central wavelength of 580 nm in the red 
(covering from 476 to 684 nm with a 5 nm gap at 580 nm, and includes the 
H$\alpha$ feature). Giraffe observed stars T01-T031 (except stars T25, T27, 
T29 and T30) with both settings HR15n with central wavelength 665 nm in the 
red (covering from 647 nm to 679 nm) and HR9 with central wavelength 525.8 nm 
(covering from 509.5 nm to 540.4 nm).

\noindent
From the abovementioned, we notice that stars T2 and T8 were observed both by 
UVES and Giraffe, providing a reliability test of these observations.

\subsubsection{UVES observations and reductions}

\noindent
With an aperture on the sky of 1 arcsec, the fibers project onto five UVES 
pixels in the dispersion direction, giving a resolving power of $\sim$47000, 
enough to determine not only the radial and rotational velocities but the 
atmosphere parameters effective temperature, surface gravity and metallicity, 
T$_{\rm eff}$, log $g$, and [Fe/H], of the stars.

\noindent
The UVES reductions were done using the UVES pipeline version 5.5.2\footnote{
http://eso.org/sci/software/pipelines}. Once individual spectra had been 
calibrated in wavelength, we corrected the spectra for the motion of the 
observatory to place them in the heliocentric reference system. Finally, we 
co-added the different exposures for each star by interpolating to a common 
wavelength array and computing the weighted mean using the errors at each 
wavelength as weights.

\subsubsection{Giraffe observations and reductions}

\noindent
The Giraffe data were reduced using the dedicated ESO Giraffe pipeline, 
version 2.15$^{10}$ and Giraffe Reflex workflow (Freudling et al. 2013), and 
calibration data provided by the ESO Science Archive Facility CalSelector 
tool. The obtained resolving power is $\sim$ 28000 for
the HR9 grating and $\sim$ 30600 for the HR15n.

\noindent
In most cases the default pipeline parameters resulted in a successful 
reduction, but in several the parameters of the flat field processing recipe 
had to be adjusted in order to achieve a successful reduction. In two cases it 
was not possible to find a set of parameters which allowed a successful 
reduction of the flat field, and in these cases the next nearest-in-time, 
successfully reduced flat was used instead.

\noindent
Each individual science observation data file was corrected for cosmic-ray 
hits using a purpose written Python script based on 
the Astro-SCRAPPY\footnote{
https://github.com/astropy/astroscrappy} (Pasquini et al. 2002) 
Python module. The cosmic-ray corrected science data files 
were processed individually with bias subtraction, flatfielding and wavelength 
calibration performed in the standard way by the pipeline and Reflex workflow. 
Subsequent reduction was then also performed in purpose written Python scripts.

\noindent
A sky spectrum was then calculated for each science data file as the median of 
all individual sky-fibre spectra available in the file, typically 15 sky 
spectra. The resulting median sky-spectrum was then subtracted from the 
spectrum extracted for each IFU fibre.

\noindent
As each star was observed with an IFU, its signal was thus distributed over 
several individual fibres, each of which is individually reduced and extracted 
by the pipeline. The S/N of each spectrum of a given IFU was then computed 
using the DERsnr 
algorithm\footnote{http://www.stecf.org/software/ASTROsoft/DER-NR/}, and the 
total signal for each star resulting from a single observation was then computed
as the sum of the signal from $N$ highest S/N spectra, which maximised the 
S/N of the summed spectrum.

\noindent
Each star in the sample was observed one to four times, depending on 
brightness. We checked that the spectra did not differ at different dates. The
final spectrum for each star was then computed as a mean of the several dates. 
The wavelength scale of the resulting summed spectra for each star was then 
corrected to the heliocentric reference system.

\subsection{Proper motions}

\noindent
To derive proper motions we used data from two HST programs collected at two 
epochs separated by almost 10 years.

\begin{table*}
\scriptsize
        \centering
        \caption{The HST data used in this work}
        \label{tab:pm0}
        \begin{tabular}{lrcc}
\\
               \hline
ACS/WFC & --- epoch 1 --- & 2003.65842-2003.66141 & \\
\hline
F502N & 4$\times$ $\sim$1100\,s & GO-9731 \\
F502N & 4$\times$  $\sim$500\,s & GO-9731 \\
F660N & 8$\times$  $\sim$500\,s & GO-9731 \\
F658N & 4$\times$ $\sim$1250\,s & GO-9731 \\
F658N & 4$\times$  $\sim$600\,s & GO-9731 \\
F550M & 4$\times$        210\,s & GO-9731 \\
\hline
WFC3/UVIS & --- epoch 2 --- & 2013.50023-2013.50666 & \\
\hline
F336W & 4$\times$  $\sim$750\,s & GO-12885 \\
F438W & 4$\times$  $\sim$270\,s & GO-12885 \\
F547M & 4$\times$  $\sim$470\,s & GO-12885 \\
F656N & 8$\times$ $\sim$1220\,s & GO-12885 \\
F658N & 4$\times$  $\sim$970\,s & GO-12885 \\
F814W & 4$\times$  $\sim$240\,s & GO-12885 \\

\hline
\hline
      \end{tabular}
      \end{table*}

\noindent
The first epoch is the data-set from GO-9731 (PI: Sankrit), and it was 
collected in August 28-29 2003 with the Wide Field Channel (WFC) of the 
Advanced Camera for Surveys (ACS). The images are in three narrow-band filters 
F502N, F660N, F658N, and in one medium-band filter, F550M.

\noindent
The second epoch is from GO-12885 (also PI: Sankrit), and it was collected in 
July 1-3 2013 with the UV-VISual Channel (UVIS) of the Wide Field Camera 3 
(WFC3). The images are in F336W, F438W, F547M, F656N, F658N, and in F814W. 
 We refer here to Table 2 for more details.

\noindent
To derive proper motions we used 28 images in each of the two 10-year apart 
epochs. In Table 2 we give a log of the used images.

\noindent
As the charge transfer efficiency (CTE) losses have a major impact on 
astrometric projects (Anderson \& Bedin 2010), in this work every single 
AC/WFC and WFC3/UVIS image employed was treated with the pixel-based 
correction for imperfect CTE developed by Anderson \& Bedin (2010). The 
improved corrections are directly included in the MAST dataproducts\footnote{
Mikulski 
Archive for Space Telescopes (MAST), at http://archive.stsci.edu/hst/search.php,
among these the $_{-}$flc exposures, which we have used}.

\noindent
Raw positions and fluxes were extracted in every image using the software and 
the spatially variable effective point spread functions (PSFs) produced by 
Anderson \& King (2006). All libraries PSFs for both ACS/WFC and WFC3/UVIS are 
publicly available\footnote{http://www.stsci.edu/∼jayander/}.

\noindent
The raw positions were then corrected for the average geometric distortion of 
these instruments using the prescriptions by Anderson (2002, 2007) for the 
ACS/WFC and by Bellini et al. (2009, 2011) for WFC3/UVIS.

\noindent
The methodology and the procedures followed to derive proper motions were 
extensively described in a previous work with similar goals (Bedin et al. 
2014). In the following we give a brief description.

\noindent
We first defined a {\it reference frame} with respect to which we will measure all 
the relative positions. This was built using the four images in filter 
WFC3/UVIS/F814W from the second epoch, as they have the best image quality and 
highest number of stars with high signal. In the reference frame we used only 
relatively bright, unsaturated, isolated stars, and with a stellar profile 
(the quality fit $q$ described in Anderson et al. 2008).

\begin{table*}
\scriptsize
       \centering

        \caption{Proper motions of the targeted stars}
        \label{tab:pm1}
        \begin{tabular}{crrrr}
\\
               \hline

NAME & dRA(mas/yr) & dDC(mas/yr)& sRA(mas/yr)&  sDC(mas/yr)\\

\hline 
  T01 &  0.96333105  & 3.42910032&   0.12081037&   0.07071602\\
  T02 & -2.33240031  & 3.86571195&  45.22749329&   54.34668660\\
  T03 &  1.65005841  &-0.00972573&   0.03339045&    0.08114443\\
  T04 & -4.85102345  &-1.31896956&   0.10271324&    0.21623776\\
  T05 &  2.78755994  & 0.98653555&   0.06256493&    0.08762565\\
  T06 & -0.14035101  & 0.80693946&   0.06580537&    0.08934278\\
  T07 &  2.22968225  & 0.63820899&   0.08311504&    0.09635703\\
  T08 & -0.64593882  & 0.48014508&  45.22705027&   54.34598682\\
  T09 &  1.15779958  & 0.38160776&   0.11226873&    0.09974361\\
  T10 & -1.17098183  &-1.93729322&   0.06072457&    0.10494832\\
  T11 &  1.07273546  & 1.85830604&   0.07653632&    0.12997579\\
  T12 & -1.84466939  &-2.21716030&   0.05498542&    0.08621992\\
  T13 &  0.82006479  & 3.46512561&   0.05292586&    0.06884338\\
  T14 & -3.85388310  & 1.59710464&   0.10770873&    0.10911011\\
  T14b&  2.19854365  &-1.00379486&   0.03274381&    0.09097165\\
  T15 & -5.93215420  & 1.39310415&   0.03482210&    0.08710643\\
  T16 &  0.29501700  & 5.03375451&   0.02070014&    0.05802598\\
  T17 &  0.31106430  & 0.16393080&   0.06300106&    0.07123875\\
  T18 & 25.06353304  &-43.80675171&  0.02033413&    0.08582691\\
  T19 & -3.14334844  &-4.36036079&   0.05860900&    0.08929045\\
  T20 & -0.81856961  & 1.70037126&   0.06403092&    0.04892372\\
  T21 & -3.56172476  &-2.59434617&   0.05555466&    0.04844967\\
  T22 &  1.63830144  &-2.74355827&   0.05628767&    0.07366438\\
  T23 &  7.54610667  &-1.28547796&   0.04192214&    0.05333274\\
  T24 & -0.54450471  & 3.25364568&   0.08436434&    0.06721759\\
  T25 &  0.16070871  &-0.27626484&  45.22705027&   54.34598682\\
  T26 & -3.39922823  & 0.75818276&   0.08496424&    0.07949148\\
  T27 &  2.52236058  & 4.58473159&   0.04709384&    0.08656006\\
  T28 &  2.09224795  & 2.28248031&   0.03391564&    0.04773037\\
  T29 &  2.81992271  &-3.87317107&   0.08391351&    0.07822889\\
  T30 &  2.37310197  & 7.78087136&  45.22739769&   54.34606942\\
  T31 & -2.61147871  &-2.23889319&   0.01148096&    0.08836027\\

\hline
\hline

\end{tabular}

\end{table*}

\noindent
The positions from all these suitable stars measured in the four F814W images 
were then transformed into a common distortion corrected reference system and 
their clipped means taken as the final positions of the reference frame. The 
field is a relatively dense one, and the final reference frame is made of over 
20 000 stars, giving a typical distance of $\sim$30 pixels to one such 
reference star. The consistency in the positions of the four F814W images
gives a positional accuracy of $\sim$0.01 pixels for the brightest 
unsaturated sources; perfectly consistent with the accuracy given in the 
Bellini et al. (2011) geometric distortion, and as achieved in other work (for 
example, Bedin et al. 2013).

\begin{figure*}
\centering
\includegraphics[width=0.75\paperwidth]{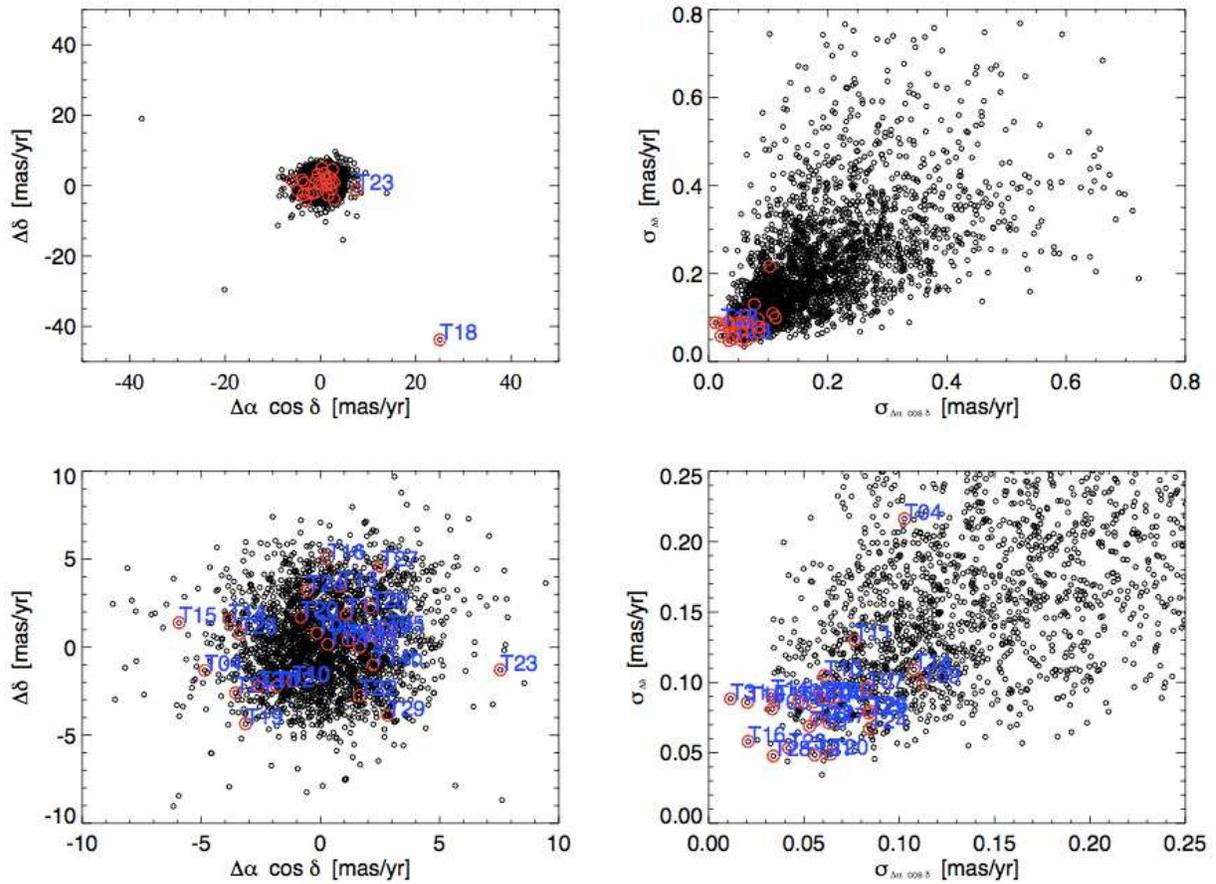}
\caption{Vector--point diagrams (left panels) of the proper motions
measured and their errors (right panels). The lower panels are 
close--ups of the upper ones. Note the very high proper motion 
of T18 and, to a lesser degree, of T23.}
\end{figure*}

\noindent
To calibrate our reference frame we need to derive the astrometric zero 
points, plate scale, orientation, and skew terms, which are needed to bring 
its coordinate systems into an absolute astrometric system. To achieve this we 
used 266 sources in common between our reference frame and the 2MASS catalog 
(Skrutskie et al. 2006).

\noindent
Hereafter our absolute coordinates will be given in equatorial coordinates at 
equinox J2000, with positions given at the reference epoch of the reference 
frame, 2013.50. Finally, we note that our absolute accuracy will be the same 
of the 2MASS catalog, about 0.2 arcsec, while our relative precisons should be 
much better than that, down to few 0.1 mas.

\noindent
As in Bedin et al. (2014), we have also produced image stacks, which provide 
often a useful representation of the astronomical scene. There are 4+6 of such 
stacks, one per filter/epoch, and they give a critical inspection of the 
region surrounding each star at any given epoch.

\noindent
We also produced trichromatic exposures for both epochs.

\noindent
As extensively explained in Bedin et al. (2014), even the best geometric 
distortion available is always an average solution. There are always 
deviations from that, typically as large as $\sim$mas. To remove these 
residual systematic errors in our geometric distortion, we used the
``bore-sight'' correction described in detail in Bedin et al. (2014), which is 
essentially a local approach (Bedin et al. 2003). We used a network of at 
least 15 stars at no more than 500 pixels from target stars, used only stars 
with consistent positions between the two epochs better than 0.75 pixels, and 
stars brigher than at least 250 photo-electrons above the local sky.

\noindent
Note that although most of the field objects are background Galactic bulge 
stars at $\sim$ 8 kpc, with an intrinsic dispersion of about $\sim$ 3 mas 
yr$^{-1}$ (e.g., Bedin et al. 2003), our proper motion precision in the 
catalogue is significantly superior to this limit. Indeed, motions of
the references local net are averaged over N reference stars, implying 
systematics lowering by the factor $\sim$ 1/$\sqrt{(N - 1)}$, with N 
typically ranging in the few hundreds and hardcoded to be at minimum 15.

\noindent
To compute proper motions, we divide the measured displacements of sources 
between epoch 2 and epoch 1, by the time baseline after having transformed the 
displacements in the astrometric reference frame provided by the sources in 
our reference frame and 2MASS.

\noindent
The proper motions are shown in  Figure 2 and in Table 3. The 
uncertainties were computed as the sum in quadrature of the r.m.s. of 
positions observed within each of the two epochs.

\begin{figure*}
\centering
\includegraphics[angle=-90,width=0.90\paperwidth]{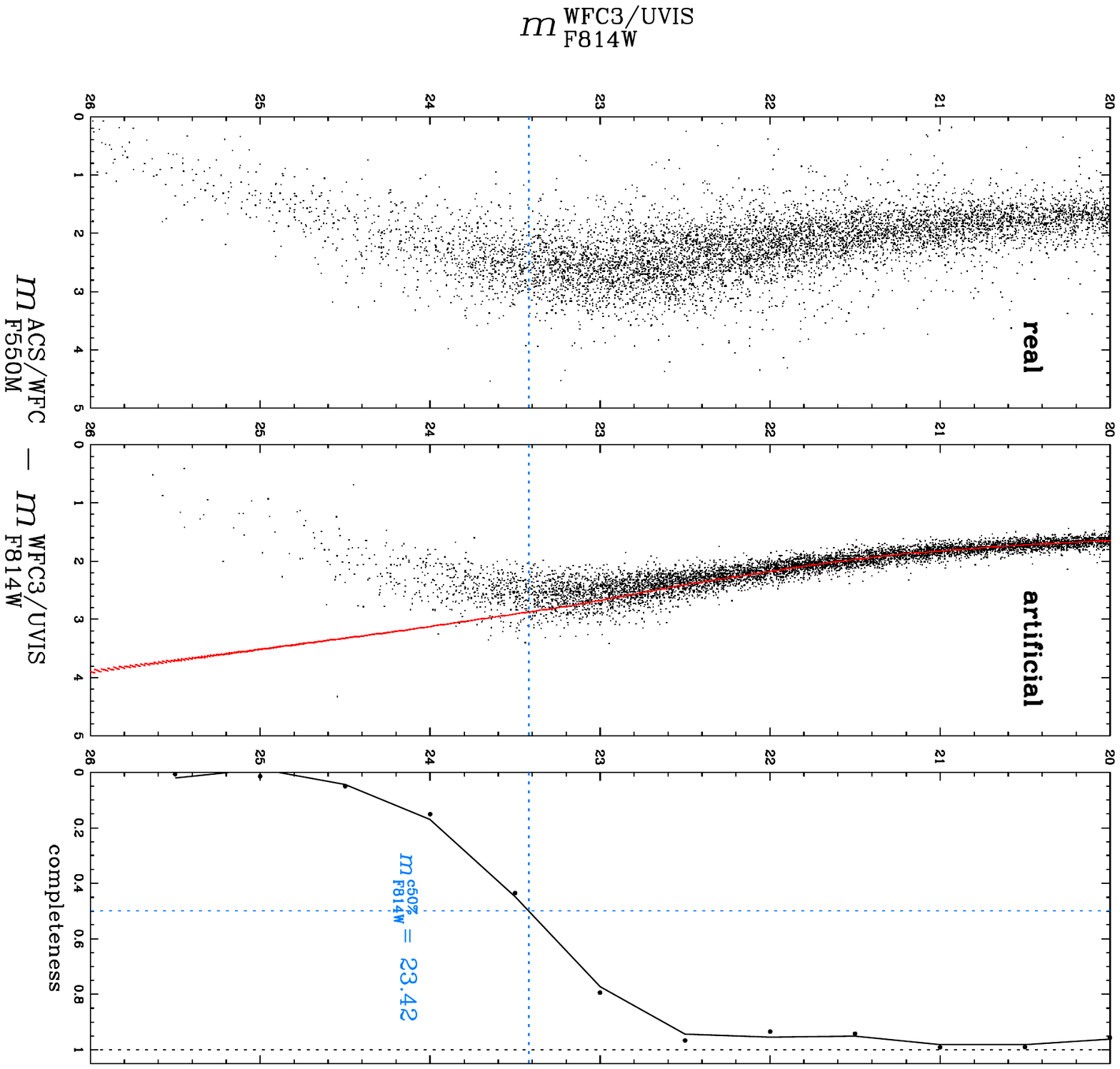}
\caption{Completeness in magnitude of the stars in the field of  SN1604.}
\end{figure*}

\noindent
To assess the completeness of the measured sample in our field we
calibrated our instrumental magnitudes and then perform artificial
star tests.

\noindent
The photometric calibration was performed only in filters
ACS/WFC/F550M (narrow $V$) and WFC3/UVIS/F814W (wide $I$) and was
obtained following the detailed procedures described in Bedin et al.\
(2005). We used the most updated zero points, which were adopted following
STScI instructions\footnote{http://www.stsci.edu/hst/wfc3/phot\_zp\_lbn}.
The calibrated CMD of the detected sources is shown in left panel of
Fig.3.

\noindent
Artificial star tests were performed as described in great detail in
Anderson et al.\ (2008) for ACS/WFC, and more recently applied to
WFC3/UVIS detectors (e.g., Bedin et al. 2015). Briefly, we built a
fiducial line of the ``main" Main-Sequence of the galactic field stars
as representative for the sources in the field (middle panel of
Fig.3), and add those sources in the individual images, which were
reduced in identical manner to the real sources.  The ratio between
the number of recovered and the inserted artificial stars at variuous
instrumental magnitude levels then provide the completness curve, which
is shown on right panel of Fig. 3.

\noindent
We can be reasonably confident that our sample is virtually complete
down to $m_{\rm F814W} \simeq 22.5$, and 50$\%$---complete down to
$\sim$23.4.  However, completeness has always a statistical value, as
individual stars can be missed for several reasons.

\bigskip

\section{Stellar parameters}

\noindent
The stellar parameters T$_{\rm eff}$, log $g$, and [Fe/H] have been determined, 
from the spectra obtained with the Giraffe spectrograph in the HR15n setup and 
from the UVES spectra in the 580nm setup, through the spectroscopic indices 
defined by Damiani et al. (2014). The method is based on a set of narrow-band 
spectral indices, and calibrated stellar parameters are derived from suitable 
combinations of these indices. Those indices sample the amplitude of the TiO 
bands, the H$\alpha$ core and wings, and temperature- and gravity-sensitive sets
of lines at several wavelength intervals. The latter are the close group of 
lines between 6490-6500  \AA, which are sensitive to gravity, and Fe I lines 
falling within the range covered by the HR15 setup, that are sensitive to 
temperature. Further gravity-sensitive features are found in the 6750-6780 
\AA\  region.

\noindent
Two global indices, $\tau$ (sensitive to temperature) and $\gamma$ 
(gravity-sensitive), are computed from the former ones. A further composite 
index, $\zeta$, measures stellar metallicity.

\noindent
Tests and calibrations of those indices have been performed (Damiani et al. 
2014), based on photometry and reference spectra from the UVES Paranal 
Observatory Project (Bagnulo et al. 2003) and the ELODIE 3.1 Library (Prugniel 
\& Soubiran 2001).

\begin{table*}
\scriptsize
       \centering

        \caption{Stellar surface parameters of the targeted stars}
        \label{tab:par}
        \begin{tabular}{lccccccccc}
\\
               \hline

Star &$v_{r}$ (km/s)&$v_{rot}\ {\rm sin}\ i$ (km/s)&T$_{\rm eff}$ (K) & T$_{\rm eff}$ err. & log $g$ & log $g$ err. & [Fe/H] &
[Fe/H] err. \\

\hline 

T01 & 34.1&14.2&4872&321&2.8&0.9&-0.4&0.3  \\
T02 & 43.9&14.2&3911&152&0.6&1.1&-1.1&0.4  \\
T03 &140.8&14.5&4415&199&2.3&0.8& 0.1&0.2  \\
T04 &  8.2&15.5&5545&251&3.8&0.6& 0.1&0.1  \\
T05 & 35.1&14.7&4053&246&0.8&1.6&-1.1&0.6  \\
T06 & 94.4&14.2&4270&196&2.1&0.8& 0.0&0.2  \\
T07 &-47.6&17.1&6243&202&3.5&0.5& 0.7&0.1  \\
T08 &-84.6&15.3&4633&159&1.9&0.5&-0.2&0.1  \\
T09 &-15.8&13.0&4890&549&4.3&1.0&-0.4&0.5  \\
T10 &-17.1&14.7&4252&191&2.0&0.8&-0.2&0.2  \\
T11 &-83.3&13.9&4000&155&1.0&0.8&-0.4&0.3  \\
T12 &-148.0&18.6&4089&192&0.9&1.0&-0.7&0.4 \\
T13 &  -4.8&13.8&4627&286&1.6&0.9& 0.2&0.2 \\
T14 & -81.3&15.0&4295&143&1.7&0.6&-0.4&0.2 \\
T14b& 142.3&12.9&4396&235&3.2&0.7&-0.0&0.3 \\
T15 & -23.2&16.4&4979&284&4.5&1.2& 0.2&0.1 \\
T16 & -89.7&14.3&4760&171&1.9&0.5&-0.3&0.1 \\
T17 &-125.8&14.9&4540&365&----&----&----&----  \\
T18 &  42.4&12.3&3777&28&4.5 &0.2&----&----  \\
T19 &   8.6&14.8&4505&351&----&----&-2.2&1.6 \\
T20 &  26.3&13.8&3874&185&1.5&1.3&-1.1&0.5 \\
T21 &  11.0&15.9&4673&189&2.0&0.6&-0.4&0.2 \\
T22 & 106.8&17.5&5164&159&2.0&0.4&-0.4&0.1 \\
T23 &  50.8&15.6&5436&336&5.3&1.6&-1.2&0.5 \\
T24 &  38.2&12.4&4237&198&3.2&0.7&-1.2&0.8 \\
T25&   45.3&20.0&4414&293&----&----&----&----  \\
T26 & -87.5&16.9&3619&316&----&----&-3.8&18.9\\
T27 & -68.3&10.7&4971&341&1.1&1.2&-1.0&0.4 \\
T28 &  51.0&14.8&4448&170&2.1&0.7&-1.0&0.3 \\
T29 &  45.4&11.3&4034&135&2.9&0.5&-0.7&0.5 \\
T30 &  54.8&11.1&4237&179&1.3&1.0&-1.3&0.4 \\
T31 & -67.5&16.1&4825&268&3.0&0.8&-0.4&0.2 \\

\hline
\hline

\end{tabular}

\end{table*}

\noindent
The method works well for stars in the approximate temperature range 3,000 K 
$\lapprox$ T$_{\rm eff}$ $\lapprox$ 9,000 K. The values of T$_{\rm eff}$, 
log $g$, and [Fe/H], with their errors, for our targeted stars, are
given in columns 4-9 of  Table 4. There were 4 stars (T17, T19, T25, and 
T26) 
whose surface gravities could not be determined in this way, due to the low 
S/N of their spectra. Their distances are thus left largely undetermined. We 
know their effective temperatures, however, and so we can estimate the 
distance depending on the luminosity classes to which they can belong.

\noindent
The resulting set of stellar parameters suggests that the stars come from a 
rather ordinary mixture of field stars (mostly giants).
 A few of the stars seem to have low [Fe/H] ($<$ -1), although with large 
errors; they are all consistent with being metal-poor giants. The radial 
velocities and the rotational $v$ sin $i$ values are very well determined. 
Radial velocities were measured from several lines in the spectra of the
targeted stars. All spectra gave a clear and narrow peak in the 
cross-correlation function (CCF) with template spectra. Even when spectral 
lines are poorly defined, the CCF may be sharply peaked, since all lines add 
up.  Radial and rotational velocities were measured from the CCF, as above, 
using a set of template spectra covering the relevant T$_{\rm eff}$ range, 
taken from the Gaia--ESO sample studied in Damiani et al. (2014); the template 
giving the highest CCF peak for each program star was used for the velocity 
determinations, an approach well-tested within the Gaia-ESO Survey. The radial 
velocities and $v$ sin $i$ values are very well determined and their errors 
(i.e. uncertainties on CCF peak center and width) lower in percentage than the 
errors in the stellar parameters. Uncertainties on radial and rotational 
velocities based on Giraffe data were carefully studied by Jackson et al. 
(2015), and found to be dependent on S/N, T$_{\rm eff}$ and $v$ sin $i$ 
(besides, obviously, of spectral resolution). For the ranges of these parameters
relevant to this work, this implies typical uncertainties of 1-2 km/s on radial
velocities, and 10-15 km/s on $v$ sin $i$.

\noindent
 T18 is  a clear outlier having very fast motion across the line of sight. 
However, this star is just a cold, M-dwarf, located at around half a kpc away only.

\section{Distances and radial velocities}

\noindent
 By comparison of the stellar parameters with the observed apparent 
magnitudes in different bands, the distances to the targeted stars are 
determined. We use, for that, the isochrones of Marigo et al. (2017)
 (see Fig. 4 for an example), which 

\begin{figure*}
\centering
\includegraphics[width=0.9\columnwidth]{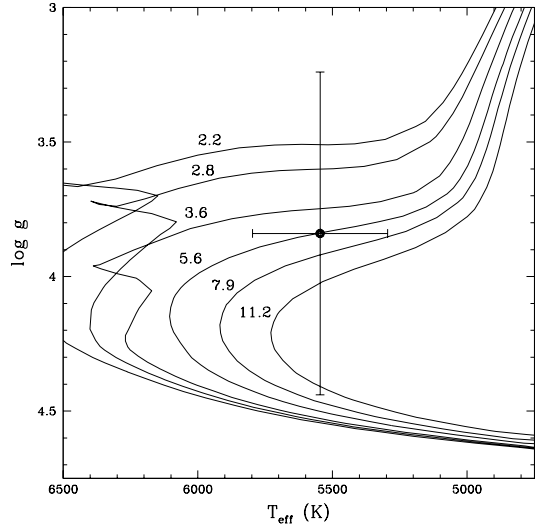}
\caption{The target star T04 plotted in the T$_{\rm eff}$--log $g$ diagram, 
superimposed on the isochrones of Marigo et al. (2017). The isochrones are
labeled in Gyrs.}
\end{figure*}

give, for each combination of the parameters T$_{\rm eff}$, log $g$, and 
[Fe/H], the absolute magnitudes M$_{V}$, M$_{R}$, M$_{J}$, M$_{H}$, and 
M$_{K}$. Apparent magnitudes of the stars in our sample are taken from the 
NOMAD catalog, and compared with the absolute magnitudes in the different 
photometric bands, taking into account the corresponding extinctions.
The results are given in Table 5. The distances indicated, with their errors, 
are weighted averages over the different bands.

\begin{table*}
\scriptsize
       \centering

        \caption{Absolute magnitudes of the targeted stars, 
                 observed apparent magnitudes, and inferred distances}
        \label{tab:dist}
        \begin{tabular}{ccccccccccccc}
\\
               \hline

Star &M$_{V}$&m$_{V}$&M$_{R}$&m$_{R}$& M$_{J}$&m$_{J}$&M$_{H}$&m$_{H}$&M$_{K}$&m$_{K}$&d (kpc)\\

\hline 

T01 &1.1$^{+1.5}_{-2.2}$&   &0.5$^{+1.4}_{-2.2}$&17.9&-0.5$^{+1.3}_{-2.1}$&15.4 &-1.0$^{+1.2}_{-2.1}$&14.8&-1.1$^{+1.3}_{-2.0}$&14.5&11.5$^{+17}_{-5.1}$\\
T02 &-4.3$^{+6.0}_{-0.7}$&   &-5.2$^{+6.4}_{-0.7}$&16.0&-7.1$^{+7.1}_{-0.6}$&  &-7.9$^{+7.3}_{-0.4}$&    &-8.1$^{+7.4}_{-0.4}$&   &$>$20\\
T03 &1.2$^{+2.2}_{-2.8}$&   &0.5$^{+2.2}_{-2.8}$&16.2&-0.8$^{+2.1}_{-2.7}$&14.3&-1.5$^{+2.0}_{-2.6}$&13.6&-1.6$^{+2.0}_{-2.6}$&13.3&8.2$^{+20}_{-2.3}$\\
T04 &3.3$^{+1.9}_{-0.7}$&  &2.9$^{+1.9}_{-0.9}$&15.9&2.2$^{+1.9}_{-1.2}$&15.1&1.7$^{+1.9}_{-1.3}$&14.5&1.6$^{+1.9}_{-1.3}$&14.5&3.3$^{+2.7}_{-1.9}$\\
T05 &-4.6$^{+4.4}_{-1.8}$&  &-5.4$^{+4.3}_{-1.8}$&17.5&-7.0$^{+4.1}_{-1.6}$&14.9&-7.7$^{+4.0}_{-1.5}$&14.1&-7.9$^{+4.0}_{-1.5}$&14.0&$>$20\\
T06 &0.8$^{+2.4}_{-2.6}$&16.3&0.1$^{+2.3}_{-2.5}$&  &-1.3$^{+2.3}_{-2.4}$&13.7&-2.0$^{+2.5}_{-2.5}$&13.0&-2.1$^{+2.3}_{-2.5}$&12.7&8.0$^{+17}_{-5.3}$\\
T07 &1.5$^{+1.9}_{-1.1}$&    &1.2$^{+1.8}_{-1.1}$&19.3&0.6$^{+1.8}_{-1.0}$&  &0.3$^{+1.8}_{-1.0}$&   &0.3$^{+1.8}_{-1.0}$&  &16$^{+11}_{-9}$\\
T08 &0.6$^{+1.1}_{-1.0}$&15.9&-0.3$^{+0.7}_{-0.7}$&12.8&-1.1$^{+0.9}_{-0.9}$&13.1&-1.9$^{+0.7}_{-0.7}$&12.3&-2.0$^{+0.7}_{-0.7}$&12.0&5.5$^{+2.0}_{-1.5}$\\
T09 &4.7$^{+4.5}_{-1.1}$&     & 4.3$^{+4.0}_{-1.1}$&    & 3.5$^{+2.7}_{-1.2}$&15.4& 3.2$^{+2.4}_{-1.3}$&14.6& 3.1$^{+2.3}_{-1.3}$&14.3&1.5$^{+2.0}_{-1.0}$\\
T10 &0.7$^{+2.3}_{-2.9}$&    &0.0$^{+2.2}_{-2.9}$&15.9&-1.4$^{+2.1}_{-2.8}$&14.9&-2.1$^{+2.2}_{-2.8}$&13.7&-2.2$^{+2.1}_{-2.8}$&13.4&12$^{+32}_{-4.0}$\\
T11 &-2.7$^{+2.5}_{-3.0}$&16.2&-3.5$^{+2.5}_{-2.9}$&13.0&-5.3$^{+2.5}_{-2.8}$&14.0&-6.0$^{+2.4}_{-2.7}$&13.2&-6.2$^{+2.4}_{-2.7}$&13.0&$>$20\\
T12 &-4.0$^{+3.1}_{-2.8}$&16.4&-4.8$^{+3.3}_{-2.4}$&    &-6.4$^{+2.7}_{-2.6}$&14.8&-7.1$^{+2.6}_{-2.8}$&14.2&-7.3$^{+2.7}_{-2.6}$&14.1&$>$20\\
T13 &-2.7$^{+2.7}_{-2.6}$&  &-3.3$^{+2.6}_{-2.4}$&17.7&-4.5$^{+2.5}_{-2.3}$&15.0&-5.1$^{+2.5}_{-2.2}$&14.2&-5.2$^{+2.5}_{-2.2}$&14.0&$>$20\\
T14 &-1.1$^{+1.6}_{-1.6}$&16.7&-1.8$^{+1.6}_{-1.6}$&13.9&-3.3$^{+1.6}_{-1.6}$&14.0&-3.9$^{+1.5}_{-1.6}$&13.2&-4.1$^{+1.6}_{-1.6}$&12.9&21$^{+25}_{-10}$\\
T14b&2.8$^{+2.3}_{-1.0}$&   &2.3$^{+2.4}_{-1.1}$&17.4&1.1$^{+2.1}_{-1.1}$&15.4&0.6$^{+2.1}_{-1.2}$&14.8&0.5$^{+2.1}_{-1.2}$&14.5&5.5$^{+4.0}_{-3.4}$  \\
T15 &5.7$^{+2.8}_{-3.2}$&16.7&5.3$^{+2.3}_{-3.3}$&13.9&4.3$^{+1.6}_{-3.3}$&  &3.9$^{+1.3}_{-3.5}$&13.2&3.8$^{+1.2}_{-3.5}$&12.9&0.6$^{+2.2}_{-0.3}$\\
T16 &-2.0$^{+2.5}_{-1.8}$&16.1&-2.6$^{+2.5}_{-1.8}$&15.3&-5.8$^{+2.4}_{-1.8}$&13.6&-4.3$^{+2.3}_{-1.8}$&12.9&-4.4$^{+2.4}_{-1.8}$&12.6&22$^{+27}_{-15}$\\
T17 &   &   &  &17.1&  &15.3&  &14.4&   &14.2&  \\
T18 &8.6$^{+0.1}_{-0.4}$&  &7.6$^{+0.2}_{-0.5}$&17.6&5.6$^{+0.4}_{-0.5}$&15.1&5.0$^{+0.4}_{-0.5}$&14.5&4.8$^{+0.4}_{-0.5}$&14.2&0.6$^{+0.2}_{-0.1}$\\
T19 &    &  &    &17.9&    &15.6&    &14.9&    &14.6&  \\
T20 &0.0$^{+3.8}_{-5.7}$&  &-0.8$^{+3.7}_{-5.7}$&17.2&-2.5$^{+3.6}_{-5.6}$&15.3&-3.3$^{+3.6}_{-5.5}$&14.5&-3.4$^{+3.5}_{-5.5}$&14.3&$>$20\\
T21 &-1.6$^{+2.6}_{-1.3}$&17.7&-2.2$^{+2.6}_{-1.3}$&16.7&-3.4$^{+2.6}_{-1.2}$&14.7&-3.9$^{+2.6}_{-1.2}$&13.9&-4.0$^{+2.6}_{-1.1}$&13.6&29$^{+20}_{-27}$\\
T22 &-2.5$^{+1.7}_{-0.9}$&16.3&-2.9$^{+1.6}_{-0.9}$&17.0&-3.9$^{+1.5}_{-0.8}$&14.1&-4.3$^{+1.5}_{-0.8}$&13.4&-4.4$^{+1.5}_{-0.7}$&13.2&28$^{+11}_{-14}$\\
T23 &9.9$^{+4.3}_{-7.8}$&16.3&8.9$^{+4.3}_{-7.1}$&15.4&6.6$^{+4.1}_{-5.5}$&14.6&6.0$^{4.0+}_{-5.1}$&14.0&5.8$^{+4.0}_{-4.9}$&13.8&0.3$^{+3.0}_{-0.2}$\\
T24 &4.3$^{+2.1}_{-2.0}$&   &3.2$^{+2.0}_{-1.9}$&17.8&2.0$^{+2.0}_{-1.8}$&15.4&1.4$^{+2.0}_{-1.8}$&14.8&1.3$^{+1.9}_{-2.2}$&14.4&3.8$^{+6.8}_{-2.2}$\\
T25 &   &   &   &   &  &12.7&  &12.0&   &11.8&  \\
T26 &   &   &   &   &  &15.6&  &15.1&   &14.8&  \\
T27 &-5.6$^{+5.4}_{-3.3}$&16.2&-6.1$^{+5.2}_{-3.2}$&  &-7.0$^{+5.0}_{-3.1}$&14.2&-7.4$^{+4.8}_{-3.6}$&13.7&-7.5$^{+4.8}_{-3.1}$&13.5&$>$20\\
T28 &0.0$^{+1.1}_{-2.2}$&  &-0.7$^{+1.1}_{-1.9}$&  &-2.0$^{+1.0}_{-2.0}$&13.9&-2.7$^{+1.1}_{-2.0}$&13.3&-2.8$^{+1.1}_{-2.0}$&13.1&13$^{+20}_{-5.0}$\\
T29 &4.1$^{+1.7}_{-1.6}$&  &3.0$^{+1.5}_{-1.6}$&  &1.7$^{+1.3}_{-1.5}$&13.8&1.1$^{+1.2}_{-1.5}$&13.0&1.0$^{+1.1}_{-1.5}$&12.8&2.1$^{+2.1}_{-0.8}$\\
T30 &-3.0$^{+3.4}_{-3.0}$&15.4&-3.7$^{+3.3}_{-3.0}$&14.9&-5.2$^{+3.3}_{-2.9}$&14.0&-5.9$^{+3.3}_{-2.8}$&13.7&-6.0$^{+3.2}_{-2.9}$&13.4&$>$20\\
T31 &2.1$^{+0.7}_{-3.0}$&   &1.4$^{+0.8}_{-2.9}$&17.8&0.3$^{+0.7}_{-2.5}$&15.8&-0.3$^{+0.8}_{-2.2}$&14.9&-0.3$^{+0.7}_{-2.2}$&14.9&9.6$^{+17}_{-2.7}$\\

\hline
\hline

\end{tabular}

\end{table*}

\noindent
Peculiar radial velocities, as referred to the average velocities of the stars 
at the same position in the Galaxy, can be one characteristic of a surviving 
companion of the SN, the excess velocity coming from the orbital motion of the 
star before the binary system is disrupted by the explosion, plus the kick 
imparted by the collision with the SN ejecta.

\begin{table}
\scriptsize
       \centering
       \caption{Radial velocities measured in this work and 
                in K14}
       \label{tab:rv}
       \begin{tabular}{lrrr}
\\ 
\hline

Star&$v_{r}$ (km/s)$^{a}$& K14& $v_{r}$ (km/s) \\

\hline

T01 & 34.1&  &          \\
T02 & 43.9&  &          \\
T03 &140.8&  &          \\
T04 &  8.2&  &          \\
T05 & 35.1&  &          \\
T06 & 94.4& P1  &  86.7 \\
T07 &-47.6& P2  &       \\
T08 &-84.6& L1  & -88.29\\
T09 &-15.8& G1  & -94.29\\
T10 &-17.1& F1  & -10.51\\
T11 &-83.3& N1  & -81.88\\
T12 &-148.0&K1  &-155.96\\
T13 &  -4.8&H1  & 177.58\\
T14 & -81.3&B1  &       \\
T14b& 142.3&B2  & 167.17\\
T15 & -23.2& D1  &-74.27\\
T16 & -89.7& E1  &-38.64\\
T17 &-125.8&     &      \\
T18 &  42.4& A1  &-69.07\\
T19 &   8.6& C1  &  7.01\\
T20 &  26.3& R1  & 41.44\\
T21 &  11.0& O1  &-10.71\\
T22 & 106.8&     &      \\
T23 &  50.8&     &      \\
T24 &  38.2&     &      \\
T25&   45.3&     &      \\
T26 & -87.5&     &      \\
T27 & -68.3&     &      \\
T28 &  51.0&     &      \\
T29 &  45.4&     &      \\
T30 &  54.8&     &      \\
T31 & -67.5& Q1  &-59.06\\

\hline
\hline
 
\end{tabular}

$^{a}$ The errors in the radial velocities are of 1--2 km s$^{-1}$.

\end{table}

\noindent
The results are given in the first and second columns in  Table 6, and 
they are compared with those of K14 for stars common to the two surveys in the 
third and fourth columns. The velocities are in the heliocentric system. There 
is good agreement for some stars but not for all of them. In our case, we have 
two or three spectra of the same star, so we can be sure about the radial 
velocities and exclude possible binarity of those stars. As an example,
we can quote the velocity of our T18, star named A in K14. We measure its 
radial velocity with an uncertainty of 1-2 km s$^{-1}$. The star moves at 42.5 
km s$^{-1}$, while K14 give -69.07 km s$^{-1}$.

\noindent
We will address the point of radial velocities and proper motions as compared 
to the kinematics of the Galaxy in Section 6.

\section{Comparison with model kinematics of the Galaxy}

\noindent
Being the surviving companion star of a SN Ia means to have a peculiar 
velocity, referred to the average velocity of the stars at the same position 
in the Galaxy, due to the orbital motion in the binary progenitor of the SN, 
plus the kick velocity caused by the impact of the SN ejecta. An estimate of 
the expected velocities, depending on the type of companion (main--sequence, 
subgiant, red giant, supergiant) was made by Canal et al. (2001), and more
recently by Han (2008). The highest peculiar velocities ($\sim$ 450 km 
s$^{-1}$ ) would correspond to main-sequence companions and the smallest ones 
($\sim$ 100 km s$^{-1}$) to red giants.
 
\noindent
As the reference for the average velocities of the stars, depending on the 
location within the Galaxy and on the stellar population considered, we adopt 
the Besan\c con model of the Galaxy (Robin et al. 2003). We have run the model 
to find the distributions of both radial velocities and proper motions in the 
direction of the center of Kepler's SNR and within the solid angle subtended 
by our search, and including all stellar populations. The same model has been 
taken as the reference in K14.

\begin{figure*}
\centering
\includegraphics[width=0.9\columnwidth]{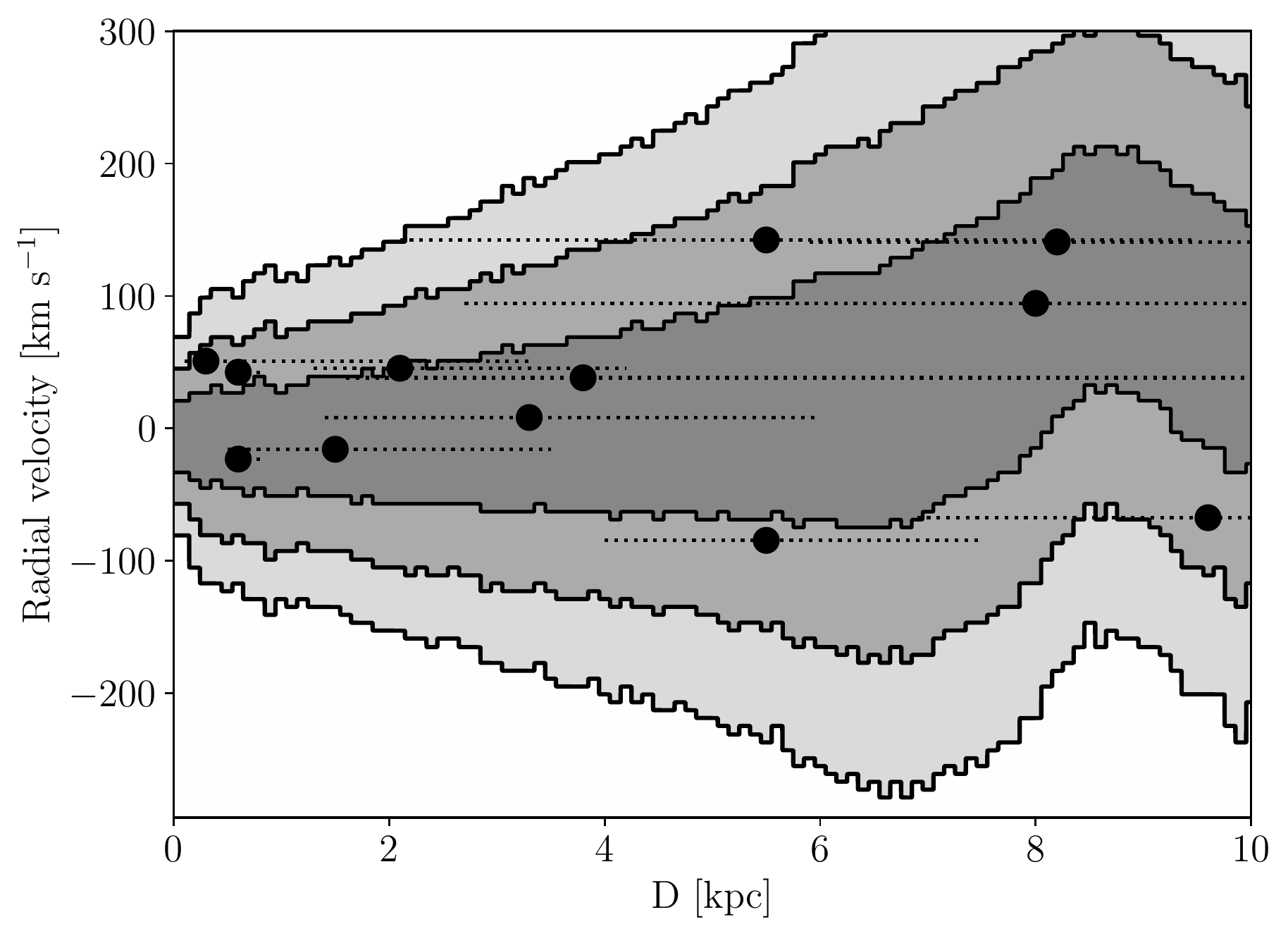}
\caption{ 
The distances and radial velocities of the targeted stars with 
$d {\lapprox 10}$ kpc, plotted over the distribution given by the Besan\c con 
model of the Galaxy. No star significantly deviates for the predicted 
distribution. We show 1 $\sigma$, 2 $\sigma$ and 3 $\sigma$ contours of the 
Galactic distribution of stars.}
\end{figure*}

\noindent
In  Figure 5, the 1, 2, and 3 $\sigma$ regions of the radial-velocity 
distribution (in the heliocentric reference system) are shown. We see that the 
average velocities first steadily increase, with positive values. That 
corresponds to the differential rotation of the Galactic disc. The dispersion 
also increases as, given the direction of the line-of-sight, we move from the 
thin to the thick disc. Then, at a distance of $\sim$7 kpc, both the slope and 
the dispersion increase when reaching the Galactic bulge, to start decreasing 
beyond $\sim$9 kpc.

\noindent
In the same Figure we compare the measured radial velocities with the 
distribution predicted by the Besan\c con model. We see that there is no star 
significantly deviating from the model distribution.

\begin{figure*}
\centering
\includegraphics[width=0.9\columnwidth]{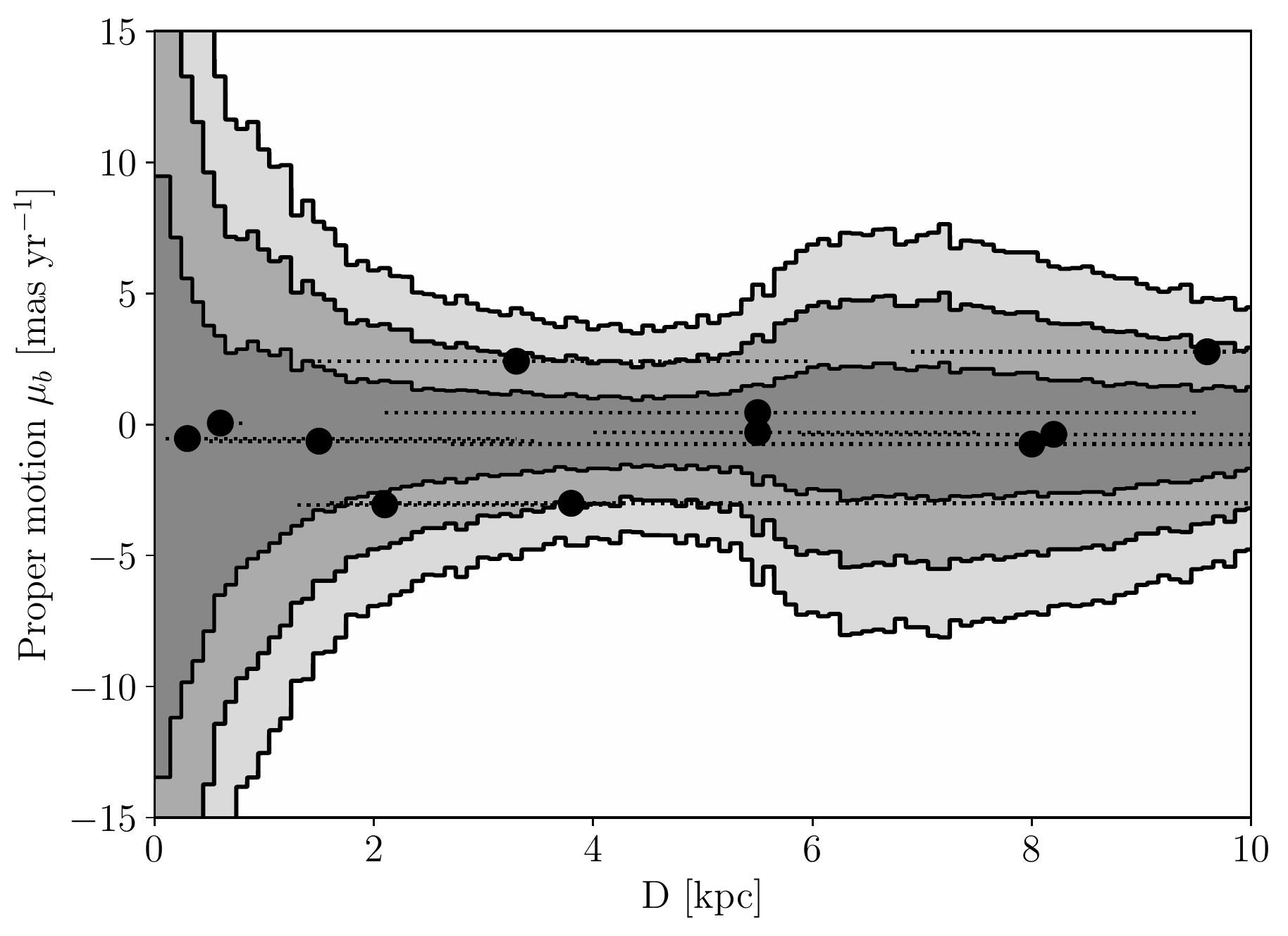}
\caption{
Same as Figure 5, for the proper motions perpendicular to the 
Galactic plane $\mu_{b}$. Star T18, at a distance of 0.4 kpc only, is not
shown here, it falling outside the scale of the plot.}
\end{figure*}

\noindent
In  Figure 6, the same is done for the proper motions perpendicular to the 
Galactic plane.  In Figure 7, the same is done for the proper motions
in Galactic latitude. 

\begin{figure*}
\centering
\includegraphics[width=0.9\columnwidth]{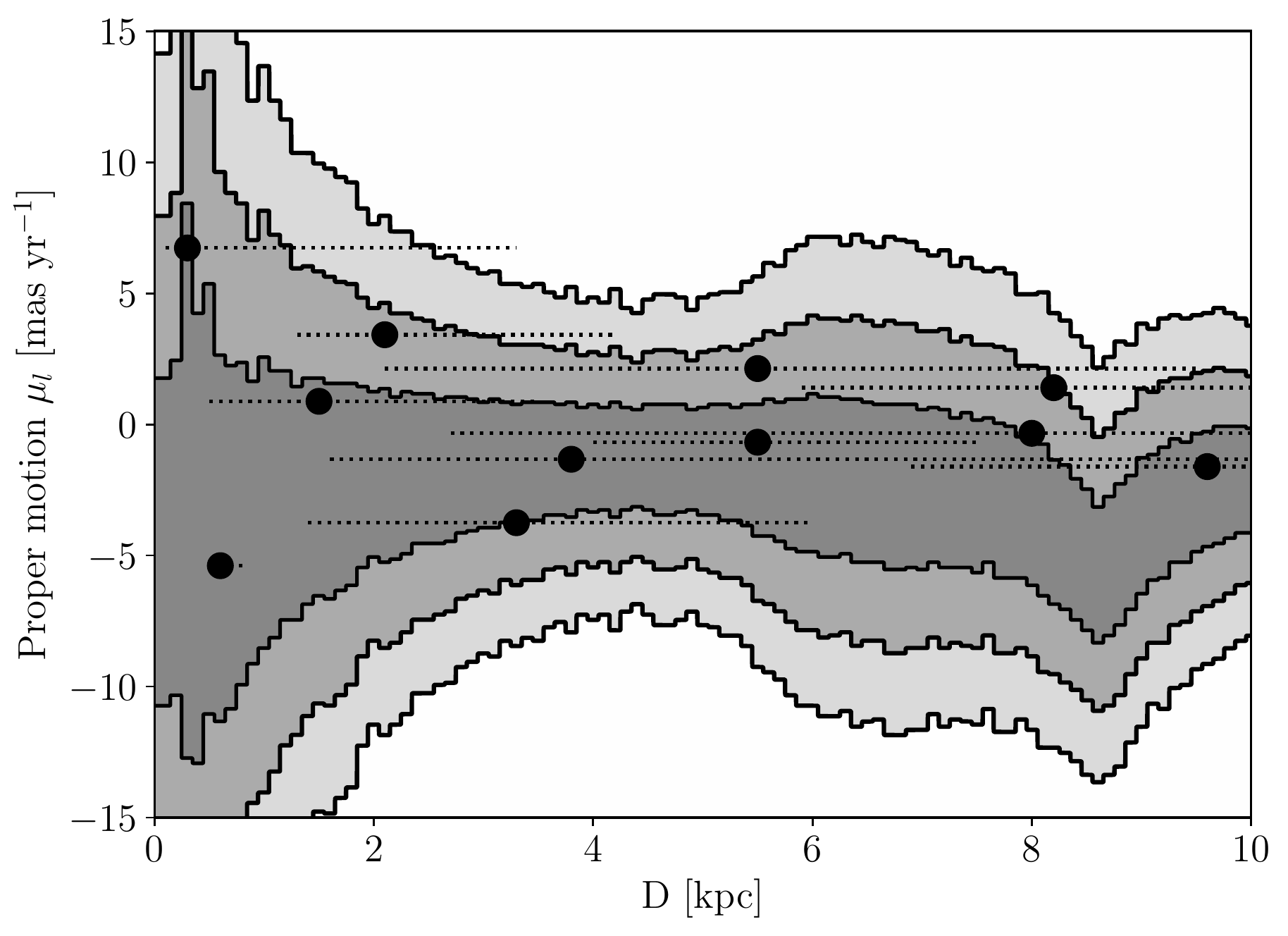}
\caption{ Same as Figure 5, for the proper motion in latitude along the 
Galactic plane $\mu_{l}$.}
\end{figure*}

\noindent
We see the same increase in the dispersion as in  Figure 5, when reaching 
the  Galactic bulge. Again, no star is a significant outlier with respect to 
the theoretical distribution. Without knowing the distances to the targets in 
 Table 5, T18 appears outstanding by its proper motion (total 
proper motion $\mu$ = 50.5 mas yr$^{-1}$), but that corresponds in fact to the 
short distance to the star, of only 0.6$^{+0.1}_{-0.2}$ kpc. For that distance, 
the velocity perpendicular to the line of sight is $v_{\rm perp}$ $\simeq$
144$^{+44}_{-48}$ km s$^{-1}$, which falls within the range of the model 
predictions. The same applies to T23, with a proper motion $\mu$ = 7.6 mas 
yr$^{-1}$, at a distance of 0.3$^{+3.0}_{-0.2}$ kpc only.  Table 7 summarizes 
the main conclusions about the stars within 10 kpc and shows the parameters in 
relation to the Galaxy model.

\noindent
K14 note that, according to Blair et al. (1991) and Sollerman et al. (2003),
Kepler's SNR has a systemic radial velocity of -180 km s$^{−1}$ . As it can be 
seen from  Figure 5, that lies between 2 $\sigma$ and 3 $\sigma$ from the 
average radial velocity of the stars at $\sim$ 5 kpc from us, in the direction 
of the SNR. K14 suggest that a possible surviving companion of the SN should
have -180 km s$^{-1}$ added to the radial velocity component of its orbital 
motion at the time of the explosion, which would make it more easily 
identifiable. The systemic velocity of the SNR, however, is that of the 
exploding WD, which is the sum of the velocity of the center of mass of the 
system plus the orbital motion of the WD at the time of the explosion. We
do not know what the velocity of the center of mass was, and thus we cannot 
just add those -180 km s$^{-1}$ .

\section{Results and discussion}

\noindent
Concerning theoretical predictions, we can compare our results with the 
observational features expected from numerical simulations. All groups that 
have simulated the impact of the ejecta of a supernova on the companion star 
(Marietta, Burrows \& Frixell 2000; Pakmor et al. 2008; Pan, Ricker \& Taam 
2012, 2013, Liu et al. 2012, 2013), find that the companion star should have 
survived the explosion and gain momentum from the disruption of the binary 
system. Their predictions vary on how luminous would the untied companion
be, on how much mass would have lost due to the impact of the supernova 
ejecta, and on how fast it would rotate and move away from the center of mass 
of the original system.

\begin{table*}
\scriptsize
       \centering

        \caption{Radial velocities, proper motions, and distances to the 
                 12 stars located at less than 10 kpc} 
        \label{tab:short}
        \begin{tabular}{lcccc}
\\
               \hline

Star &$v_{r}$ (km/s)$^{a}$&$\mu_{l}$ (mas/yr)&$\mu_{b}$ (mas/yr)&d (kpc)\\

\hline 

T03 &    140.8&  1.40$\pm$0.00&  -0.38$\pm$0.08&8.2$^{+20}_{-2.3}$\\
T04 &      8.2& -3.75$\pm$0.03&   2.41$\pm$0.24&3.3$^{+2.7}_{-1.9}$\\
T06 &     94.4& -0.34$\pm$0.03&  -0.75$\pm$0.10&8.0$^{+17}_{-5.3}$\\
T08 &    -84.6& -0.68$\pm$23.70& -0.31$\pm$63.01&5.5$^{+2.0}_{-1.5}$\\ 
T09 &    -15.8&  0.88$\pm$0.07&  -0.64$\pm$0.11&1.5$^{+2.0}_{-1.0}$\\
T14b&    142.3&  2.13$\pm$0.00&   0.45$\pm$0.09&5.5$^{+4.0}_{-3.4}$\\
T15 &    -23.2& -5.39$\pm$0.00&   0.06$\pm$0.09&0.6$^{+2.2}_{-0.3}$\\
T18 &     42.4& 32.94$\pm$0.01&  36.31$\pm$0.09&0.6$^{+0.2}_{-0.1}$\\
T23 &     50.8&  6.73$\pm$0.02&  -0.53$\pm$0.06&0.3$^{+3.0}_{-0.2}$\\ 
T24 &     38.2& -1.33$\pm$0.05&  -3.00$\pm$0.09&3.8$^{+6.8}_{-2.2}$\\  
T29 &     45.4&  3.42$\pm$0.05&   3.06$\pm$0.10&2.1$^{+2.1}_{-0.8}$\\
T31 &    -67.5& -1.61$\pm$0.02&   2.77$\pm$0.09&9.6$^{+17}_{-2.7}$\\

\hline
\hline

\end{tabular}

$^{a}$ The errors in the radial velocities are of 1--2 km/s.

\end{table*}

\noindent
Let us start with the rotational velocities of the post-explosion companions. 
Liu et al. (2013) did binary population synthesis after performing 3D 
hydrodynamic simulations of the impact of the ejecta on a main sequence star 
with different orbital periods and separations from the exploding WD. They 
obtained the expected distribution of rotational velocities for the surviving 
companion. It leaves room for a wide range in this parameter, unlike previous 
assumptions that the post-impacted star should have very high velocities.
Pan, Ricker \& Taam (2012) also found that angular momentum of the companion 
would have been lost with the stripped material. In the case of the stars 
studied in the survey for the companion of SN 1604, all of them have 
rotational velocities lower than 20 km s$^{-1}$. This is not uniquely 
interpreted as a sign of the absence of companions, though.

\noindent
Concerning the luminosity discussion, there are some differences in the way 
the surviving companions from the supernova explosion would be. Podsiadlowski 
(2003) found that, for a subgiant companion, the object $\sim$ 400 years after 
the explosion might be significantly overluminous or underluminous relative to 
its pre-SN luminosity, that depending on the amount of heating and the amount 
of mass stripped by the impact of the SN ejecta. More recently Shappee, 
Kochanek \& Stanek (2013) have also followed the evolution of luminosity
for years after the impact of the ejecta on the companion. The models in which 
there is mass loss rise in temperature and luminosity peaking at 10$^{4}$ 
L$_{\odot}$ to start cooling and dimming down to 10 L$_{\odot}$ some 10$^{4}$ yr 
after the explosion. Around 500 days after explosion the companion luminosity 
is 10$^{3}$ L$_{\odot}$. Pan, Ricker \& Taam (2012, 2013, 2014) found lower
luminosities for the companions than these previous authors. They found 
luminosities of the order of 10 L$_{\odot}$ for the companions, several 
hundred days after the explosion. It is interesting to see that they predict, 
for the surviving companions, effective temperatures, T$_{\rm eff}$ in the 
range 5000-9000 K. This allows to discard possible candidates below 5000 K in
our sample. Only  four stars in our sample are at T$_{\rm eff}$ higher than 
5000 K (see Table 4 and Table 7). T04 has 5545 $\pm$ 251 K and 
log $g$ = 3.8 $\pm$ 0.6. The 
distance is uncertain but consistent with that of Kepler's SN: a distance of 
3.3$^{+2.7}_{-1.9}$ kpc. The heliocentric velocity, however, is
$v$ = 8.2 km s$^{-1}$ and the proper motion, very moderate, $\mu_{l}$ = 3.75 
$\pm$ 0.03 mas yr$^{-1}$ and $\mu_{b}$ = 7.41 $\pm$ 0.24 mas yr$^{-1}$. So, it 
is within the expectations of the Besan\c con model. There is only one more 
target at a distance compatible with Kepler's SN explosion
 and {$T_{\rm eff}$) higher than 5000K. This  
target is T23, with a T$_{\rm eff}$ of 5436 $\pm$ 336 K and log $g$ = 5.3 $\pm$ 
1.6,  with distance of 0.3$^{+3.0}_{-0.2}$ kpc. The radial veocity is 50.8 
km s$^{-1}$ and the proper motion is $\mu_{l}$ = 6.73 $\pm$ 0.02 mas yr$^{-1}$ 
and $\mu_{b}$ = -0.53 $\pm$ 0.06 mas yr$^{-1}$. 

\noindent
At a distance compatible with the Kepler SN there are no main sequence stars 
in our sample (let us recall that we go down to 2.6 L$_{\odot}$). There 
are 4 subgiants and 2 giants at distances compatible with the Kepler distance. 
Their stellar parameters, radial velocities, rotational velocities and proper 
motions are within what is expected of a sample field at the Kepler position 
in our Galaxy.

\noindent
Target 18 
(star A in K14) has a proper motion of $\mu_{l}$ = 32.94 $\pm$ 
0.09 mas yr$^{-1}$ and $\mu_{b}$ = 36.31 $\pm$ 0.01 mas yr$^{-1}$. This is 
a clear outlier in proper motion and it is crucial to determine its stellar 
parameters and distance. It turns out that the star is at 0.6$^{+0.2}_{-0.1}$ 
kpc only. It is an M star belonging to the main sequence.

\noindent
Overall, there are no stars showing any peculiarity. All of them have 
rotational velocities around 10-20 km s$^{-1}$ or less, since Giraffe HR15n is 
not able to measure values lower than that. Their radial velocities are within 
those expected for field stars. 

\noindent
The predictions by Pan, Ricker \& Taam (2012, 2013) were that 
400 yrs after the SN Ia explosion, the luminosities of the 
companion stars would still be 10 times higher than those before receiving the 
impact of the ejecta of the SN. They have extended those predictions to
main sequence (MS) companion masses down to 0.656 M$_{\odot}$ and He WDs down 
to 0.697 M$_{\odot}$ (Pan, Ricker \& Taam 2014). We have gone below the 
luminosities predicted for surviving companions of the kind examined by these 
authors and the predicted T$_{\rm eff}$ are higher than those found in our 
sample. They  have calculated the post-impact evolution of MS 
companions and He--WD companions of very low mass at the time of the explosion, 
and also the post-impact evolution of these companions. 

\noindent
The He WDs at the time of the explosion (Table 1 in Pan, Ricker \& Taam 2014) 
have runaway velocities within the range 490-730 km s$^{-1}$, which would 
correspond, for purely transversal motions at a distance of 5 kpc, to proper 
motions $\mu$ between 21 and 31 mas yr$^{-1}$ or, if assumed to make a 45$^{o}$
angle with the line of sight, to radial velocities between 350 and 516 km 
s$^{-1}$ and proper motions between 15 and 22 mas yr$^{-1}$ Those proper 
motions would have been detected by the {\it HST} astrometry, even for objects 
fainter than our targets down to m$_{F814W}$ $\sim$ 22.5 mag.

\noindent
There are several channels through which WDs could be surviving companions of 
SN Ia explosions, apart from the He--WDs companion abovementioned. One is 
dynamically stable accretion on a CO WD from a He-WD or from a lower-mass CO  
WD (Shen \& Schwab 2017). In that case, a He-shell detonation could induce a 
core explosion (Shen \& Bildsten 2014). The mass-donor WD might survive. One 
salient characteristic of those companions is that, due to their extreme 
closeness to the exploding WD and to their strong gravitational fields, they 
should capture part of the radioactive material ($^{56}$Ni) produced by the SN.

\noindent
Shen \& Schwab (2017) study the effects of the decays of $^{56}$Ni to $^{56}$Co 
and of $^{56}$Co to $^{56}$Fe, for different masses of captured material by WDs 
of masses between 0.3 M$_{\odot}$ and 0.9 M$_{\odot}$. The decays, in the 
physical conditions prevailing at the surfaces of those WDs, drive persistent
winds and produce residual luminosities that, 400 yr after the explosion, are 
higher than $\sim$10 L$_{\odot}$ in all cases (see Fig. 4 in Shen \& Schwab 
2017). Furthermore, the surviving WDs should be running away from the site of 
the explosion at velocities $\sim$ 1500-2000 km s$^{-1}$. A search for such WD 
companions has recently been made by Kerzendorf et al. (2018), in the
central region of the remnant of SN 1006, with negative result. We have not 
detected faint hot surviving WDs moving at high 
speed. We are at larger distance than SN 1006 and the exploration 
does not go so deep, though (see our completeness discussion).  

\noindent
Another possible channel producing a surviving WD companion is the spin-up, 
spin-down model (Justham 2011; Di Stefano, Voss \& Claeys 2011): the WD, spun 
up by mass accretion from the companion star, can grown beyond the 
Chandrasekhar mass; then, when the accretion ceases, it has to lose angular
momentum before reaching the point of explosion. During this last time 
interval, the companion might have evolved past the AGB stage and become a 
cool WD. The time scale for spin down is hard to be determined theoretically,
but Meng \& Podsiadlowski (2013) empirically obtain an upper limit of a few 
10$^{7}$ yr, for progenitor systems that contain a RG donor and for which 
circumstellar material has been detected. We must note, however, that the 
spin-up, spin-down model should mostly produce super-Chandrasekhar explosions, 
since there is nothing there to tell the system to stop mass transfer just 
when the WD has reached that mass. In the case of Kepler's SN, reconstruction 
of its light curve (Ruiz-Lapuente 2017) clearly indicates that the SN was in
no way overluminous.

\noindent
From all the preceding, we can exclude MS, subgiants, giants,  and 
up to certain extend stars below the solar luminosity.

\noindent
As an interesting point, no one has yet attempted to calculate how much and 
for how long the impact of the SN Ia ejecta would affect the luminosity of a 
WD companion in the spin-up, spin-down case. One can not just assume 
that the WD 
would be cold and dim and remain so after the explosion. This has not been 
proved by any hydrodynamic simulation. It has only been done for closer pairs 
of WDs, as mentioned above. The typical separation between the two WDs, at the 
time of the explosion, should be larger that in the cases considered by Pan, 
Ricker \& Taam (2014) and Shen \& Schwab (2017), so less radioactive
material would be captured and the runaway velocities would also be smaller, 
but the narrowing of the orbit by the emission of gravitational waves during 
the cooling stage of the companion WD might not be negligible, and the loss of 
angular momentum by the system during a likely common-envelope episode 
preceding the formation of the detached WD-WD system would also have 
considerably narrowed, previously, the separation between the two objects. We 
encourage to do these hydrodynamical calculations.

\section{Conclusions}

\bigskip

\noindent
 We present a study that includes the first detailed stellar parameters:
Teff, , log g, v$_{rot}$; as well as  accurate radial velocities
of the stars, and proper motions, using the {\it HST}, of possible 
companions of Kepler's SN within 20 \% of the remnant center. This last part 
of the research is very important, since one 
does not know whether the peculiar velocities expected for the surviving 
companion will mostly be along the line of sight or perpendicular to it. No 
attempt to measure the proper motions of the stars in the core of Kepler's SNR 
had ever been made before.

\noindent
We have determined  luminosities and distances to the 
candidate companions of Kepler's SN. Any companion would have luminosities 
above two times the solar luminosity which is the lowest luminosity of our 
sample. The radius of our search is 24 arcsecs, that is 20\% of the average 
radius of the SNR. Our stars correspond to stellar parameters and velocities 
consistent with being from a mixture of stellar populations in the direction 
of the Kepler SNR.

\noindent
From our study, we conclude that the single- degenerate scenario is disfavored 
in the case of Kepler's supernova. The idea that Kepler's SN could come from 
the merging of two stars within a common envelope seems plausible. It would 
explain why the SN is surrounded by a large circumstellar medium (CSM). The 
idea of the core-degenerate scenario (Kashi \& Soker 2011), that an already 
existing WD and a degenerate RG core merge inside an AGB envelope, appears 
very likely in this case.

\noindent
This analysis makes relevant intensive studies to detect surviving companions 
in very nearby SNeIa remnants. There are many good cases for study in our 
Galaxy and in the nearby ones. There are cases in our Galaxy  far away 
enough so that Gaia can not make proper motion estimates, the stars being too 
dim. The {\it HST} plays a key role here. In addition, telescope
time in 10m-class telescopes and in the coming generation of large telescopes 
with high resolution spectrographs is the key to determine the nature of the 
surviving companions of Type Ia SNe.

\noindent
 As mentioned in the discussion, more hydrodynamical simulations are needed
to compare predictions with  observational results.

\bigskip

\bigskip

\noindent
{Acknowldegments}

\noindent
Based on observations made with ESO Telescopes at the La Silla Paranal
Observatory under programme ID 093.D-0384(A). Based on archival images from 
the {\it HST} programs GO-9731 and GO-12885. The scientific results reported 
in this article make use of observations by Chandra X-ray Observatory and 
published previously in cited articles. This research was supported by the 
Munich Institute for Astro- and Particle Physics (MIAPP) of the DFG cluster of 
excellence ``Origin and Structure of the Universe''. P.R-L is supported by
AYA2015-67854-P from the Ministry of Industry, Science and Innovation of Spain 
and the FEDER funds. J.I.G.H. acknowledges financial support from the Spanish 
MINECO under the 2013 Ramon y Cajal program MINECO RyC-2013-14875, and also 
from the Spanish Ministry Project MINECO AYA2014-56359-P. L.G. was supported 
in part by the US National Science Foundation under Grant AST-1311862.
We thank the referee, Wolfgang Kerzendorf, for his very  useful  
report.   

\bigskip

}

\end{document}